\documentclass[aip,preprint]{revtex4-1}
\usepackage{amsmath,amssymb,graphicx}

\begin{document}

\title{ Realization of an all optical exciton-polariton router }

\author{F\'elix Marsault}
\affiliation{Laboratoire de Photonique et de Nanostructures,
LPN/CNRS, Route de Nozay, 91460 Marcoussis, France}
\author{Hai Son Nguyen}
\affiliation{Laboratoire de Photonique et de Nanostructures,
LPN/CNRS, Route de Nozay, 91460 Marcoussis, France}
\author{Dimitrii Tanese}
\affiliation{Laboratoire de Photonique et de Nanostructures,
LPN/CNRS, Route de Nozay, 91460 Marcoussis, France}
\author{Aristide Lema\^{\i}tre}
\affiliation{Laboratoire de Photonique et de Nanostructures,
LPN/CNRS, Route de Nozay, 91460 Marcoussis, France}
\author{Elisabeth Galopin}
\affiliation{Laboratoire de Photonique et de Nanostructures,
LPN/CNRS, Route de Nozay, 91460 Marcoussis, France}
\author{Isabelle Sagnes}
\affiliation{Laboratoire de Photonique et de Nanostructures,
LPN/CNRS, Route de Nozay, 91460 Marcoussis, France}
\author{Alberto Amo}
\affiliation{Laboratoire de Photonique et de Nanostructures,
LPN/CNRS, Route de Nozay, 91460 Marcoussis, France}
\author{Jacqueline Bloch}\email{jacqueline.bloch@lpn.cnrs.fr}
\affiliation{Laboratoire de Photonique et de Nanostructures,
LPN/CNRS, Route de Nozay, 91460 Marcoussis, France}
\affiliation{ Physics Department, Ecole Polytechnique, F-91128 Palaiseau Cedex, France}

\date{\today}

\begin{abstract}
 We report on the experimental realization of an all optical router for exciton-polaritons. This device is based on the design proposed by H. Flayac and I.G. Savenko [Applied Physics Letters 103, 201105 (2013)], in which a zero-dimensional island is connected through tunnel barriers to two periodically modulated wires of different periods. Selective transmission of polaritons  injected in the island, into either of the two wires, is achieved by tuning the energy of the island state across the band structure of the modulated wires. We demonstrate routing of ps polariton pulses using an optical control beam which controls the energy of the island quantum states thanks to polariton-exciton interactions.
\end{abstract}

\maketitle 

Exciton-polaritons in semiconductor microcavities have attracted growing interest in the past few years, not only because of their fascinating fundamental properties (for a review see Refs.~\onlinecite{Byrnes2014,Carusotto2013}) but also for their potential interest for applications. Polaritons arise from the strong coupling between excitons confined in quantum wells and the optical mode of a microcavity~\cite{Weisbuch1992}. These hybrid light-matter quasi-particles present a Kerr-like non linearity which is amongst the largest in optical systems. Using this giant non-linearity, polaritonic parametric oscillator \cite{Stevenson2000,Diederichs2006} and amplifier \cite{Savvidis2000,Saba2001}, low threshold bistability \cite{Baas2003} and multistability \cite{Paraiso2010} have been demonstrated. Semiconductor microcavities are now foreseen as a promising platform for integrated photonics, where coherent emission, optical guiding and non-linearity can be combined within the same chip. Elementary building blocks for optical processing have been realized such as spin-switch \cite{Amo2010a}, spin-memory \cite{Cerna2013}, optical transistors \cite{Gao2012,Ballarini2013a}, logic gates \cite{Ballarini2013a,Anton2013} and resonant tunneling diodes \cite{Nguyen2013}. More advanced  architectures for polaritonic circuits are now proposed \cite{Liew2008a,Espinosa-Ortega2013}.

Since polaritons have a very low effective mass, typically 5 orders of magnitude smaller than the bare exciton, a lateral optical confinement on a micron scale can produce strong enough confinement to lower the polariton dimensionality, design circuits or engineer their  band-structure. A variety of  techniques have been proposed and implemented recently to pattern microcavities : deposition of metallic layers on top of the cavity \cite{Lai2007}, creation of a potential by optical means \cite{Amo2010} or using surface acoustic waves \cite{Cerda-Mendez2010}. Stronger confinement can be achieved by patterning mesas during the cavity growth \cite{Kaitouni2006,Winkler2015}, by using a photonic crystal as top mirror \cite{Zhang2014} or by directly etching the microcavity \cite{Bloch1997}.

Recently, deep etching was implemented to realize polaritonic devices such as an optically controlled Mach Zehnder interferometer \cite{Sturm2014}, or a resonant tunneling diode \cite{Nguyen2013} where the energy of a discrete state is optically tuned to modulate the device transmission. Strong modifications of the polariton band structure have been reported when laterally modulating the width of a one dimensional cavity: regular minigaps and minibands are formed in the case of a periodic modulation and condensation occurs in solitonic gap states\cite{Tanese2013}.

Inspired by these recent realizations and combining the band structure of  periodically modulated wires\cite{Tanese2013} with the operating concept of the polariton diode\cite{Nguyen2013}, H. Flayac and I.G. Savenko \cite{Flayac2013} recently proposed a design to realize a polariton router. It is based on an island connected to two modulated wires through tunnel barriers. Polaritons can tunnel through the barriers and propagate in the wires if their energy matches a wire miniband, whereas on the contrary they are blocked within the island if their energy lies within a forbidden gap. To implement the routing functionality, an asymmetric design is proposed: the engineered periods of the two wires are different so that it is possible to tune the island state into a forbidden gap of one of the wires while being resonant with a mini-band of the other, and vice versa. Thus the output wire for polaritons can be selected by tuning properly the island state energy.

In this letter, we report on the realization of this polariton router implementing the design proposed by H. Flayac and I.G. Savenko \cite{Flayac2013}. Etching a high quality factor microcavity, we fabricate a zero-dimensional (0D) island with confined polariton states coupled to two periodically modulated wires. Far field emission of the wires shows different band structures for the right and left wires with different energy gaps. Polaritons are selectively injected within the island and their transmission in each wire is monitored. As we increase the excitation power, the lowest energy state of the island is continuously blueshifted because of polariton exciton interactions. As a result, the state is progressively swept across the band structure of each modulated wire. Suppression of the transmission into the modulated wire is observed when the island state enters a forbidden energy gap. This occurs successively in one and then in the other wire. We use this control of the tunnel coupling to implement a polariton router. We demonstrate routing of polariton pulses into the wire by means of a cw non-resonant gate beam which tunes the energy of the island.

\begin{figure}[!ht]
\centerline{\includegraphics[width=\columnwidth]{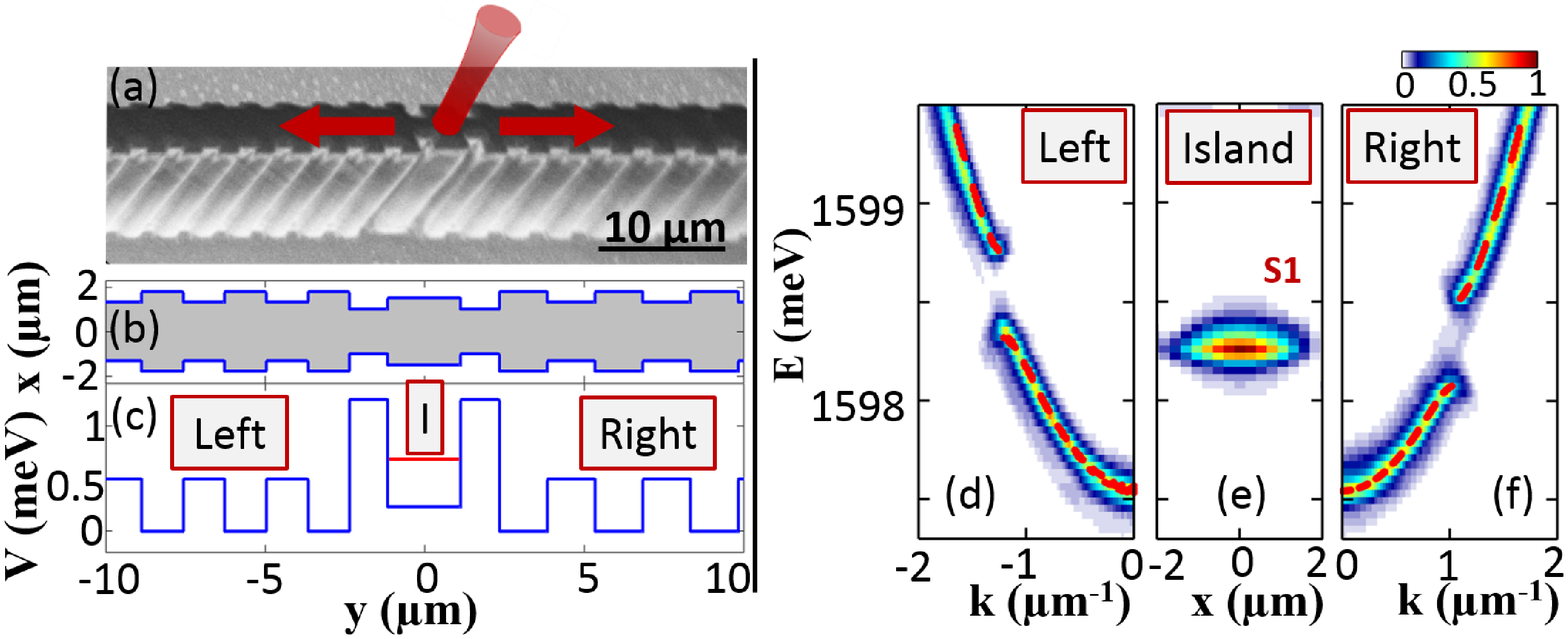}}
\caption{(a) Scanning electron microscopy image of the device. In red, the excitation laser and polariton flows are schematically shown;
(b) Schematic top view of the device;
(c) (blue) Simulated energy potential along the wire and  island regions and (red) energy of the first confined state S1 within the island;
(d)-(f) Spectrally resolved far field emission of the (d) left and (f) right wire. Red dotted lines are fits of the dispersions solving a 2D Schr\"{o}dinger equation for polaritons;
(e) Spectrally and spatially resolved emission of the island. Experiments are performed using a $50\,\mu W$ excitation power.
}
\label{Fig1}
\end{figure}

The sample was grown by molecular beam epitaxy and consists in a $ \lambda / 2 \ \mathrm{Ga}_{0.05}\mathrm{Al}_{0.95}\mathrm{As} $ layer surrounded by two $\mathrm{Ga}_{0.8}\mathrm{Al}_{0.2}\mathrm{As} / \mathrm{Ga}_{0.05}\mathrm{Al}_{0.95}\mathrm{As} $ Bragg mirrors with $28$ and $40$ pairs in the top/bottom mirrors, respectively.  Twelve GaAs quantum wells of width $7 \,\mathrm{nm}$ are inserted at the anti-nodes of the electromagnetic field, resulting in a $15 \,\mathrm{meV}$ Rabi splitting.
The polariton router is fabricated using electron beam lithography and dry etching [Fig.~\ref{Fig1}(a)]. It is made of a 0D island ($2.3 \,\mu \mathrm{m}$ long and $2.8 \,\mu \mathrm{m}$ wide) surrounded by two constrictions ($1.2 \,\mu \mathrm{m}$ long and $2 \,\mu \mathrm{m}$ wide). These constrictions define tunnel barriers connecting  the island to two $150 \,\mu \mathrm{m}$ long periodically modulated wires with a period of $2.6 \,\mu \mathrm{m}$ ($3 \,\mu \mathrm{m}$) for the left (right) wire. The wire transverse dimension is modulated between $2.6 \,\mu \mathrm{m}$ and $3.5 \,\mu \mathrm{m}$ for both wires [Fig.~\ref{Fig1}(b)]. The resulting potential profile along the wire is shown on Fig.~\ref{Fig1}(c) together with the first confined polariton mode in the island.

We perform microphotoluminescence experiments maintaining the sample at $10 \,\mathrm{K}$. The sample is excited non-resonantly with a cw mono-mode Ti:sapphire laser tuned typically $100 \,\mathrm{meV}$ above the island lowest energy state. For dynamical experiments, an additional pulsed Ti:sapphire laser delivering $3 \,\mathrm{ps}$ pulses with a $80 \, \mathrm{MHz}$ repetition rate is used to excite resonantly the island quantum state. Both lasers are linearly polarized parallel to the wires and focused onto a $2 \,\mu \mathrm{m}$ spot with a microscope objective ($\mathrm{NA}=0.65$). The sample emission is collected  with the same objective, filtered through a polarizer parallel to the wire, and focused on the slit (parallel to the wire) of a spectrometer which is coupled to a CCD camera or to a streak camera (for time resolved measurements).
For the microwire probed in the experiments reported below, the exciton-photon detuning $\delta = E_{cav}-E_{exc}$  amounts to $-3.6\, \mathrm{meV}$.

We first characterize the microstructure by exciting only the island or one of the wires with a non resonant beam at low power and we monitor the polariton states in each region using a spatial filtering. Such etched microstructures present a polarization splitting and two subspaces of polariton eigenstates with orthogonal linear polarization\cite{Sturm2014}. Here we restrict our measurements to one of these subspaces by measuring one polarization. The polariton band structure in reciprocal space, obtained by imaging the far field emission, is shown on Fig.~\ref{Fig1}(d) and Fig.~\ref{Fig1}(f) for the left and right wires, respectively. The formation of minibands and the opening of an energy band gap because of the periodic potential is clearly observed. We measure on both wires a gap width of  $500\, \mu \mathrm{eV}$. Because of the different periods in the left and right wire, the center of the gaps are at different energies, $300\,\mu \mathrm{eV}$ apart from each other, a key feature for the operation of the polariton router. The island emission pattern in real space is displayed in Fig.~\ref{Fig1}(e) and shows a well defined discrete polariton state corresponding to the lowest energy polariton state S1 confined within the island. Its measured linewidth is $\Gamma = 120 \, \mu \mathrm{eV}$, thus significantly smaller than the bandgap spectral width of both wires\cite{Largeurderaie}.

\begin{figure*}[!htbp]
	\centerline{\includegraphics[width=\columnwidth]{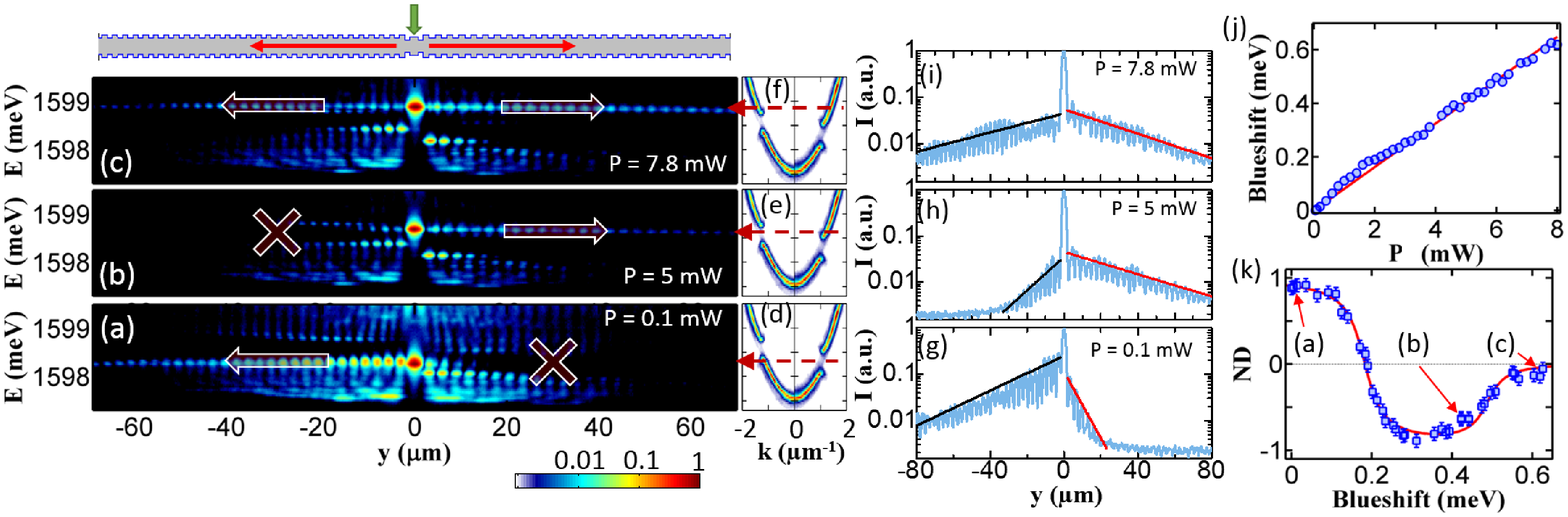}}
	\caption{(a-c) Spectrally and spatially resolved emission measured for an excitation power (a) $P = 0.1 \,\mathrm{mW}$, (b) $5 \,\mathrm{mW}$, and (c) $7.8 \,\mathrm{mW}$;
		(d-f) Spectrally resolved far field emission of the left wire (left part) and of the right wire (right part) for the same three values of \textit{P};
		(g-i) Integrated intensity spatial profile measured at the island energy  extracted from (a), (b) and (c) respectively. Black and red lines are guides to the eye;
		(j) S1 state blueshift as a function of \textit{P};
		(k) Measured $ND$ as a function of S1 blueshift. Red line: \textit{ND} calculated from the dispersions shown in Fig.~1.
	}
	\label{Fig2}
\end{figure*}

We now demonstrate how we can sweep the energy $E1$ of the island state S1 across the wire band structure, and thus modify its tunnel coupling to the modulated wires.

We use non resonant cw excitation to excite locally the island. Both the island states and higher energy excitonic states, forming the so-called exciton reservoir, are populated in this excitation scheme. As the excitation power is increased, repulsive polariton-exciton interactions induce a continuous blueshift of the polariton state within the island \cite{Wouters2008,Wertz2010}. Thanks to this strong optical non-linearity of polaritons, we can tune $E1$ in the present experiment by $0.6 \,\mathrm{meV}$ as shown in Fig.~\ref{Fig2}(j).

Spatially resolved emission measured along both wires on each side of the island is shown in Figs.~\ref{Fig2}(a,b,c), for various excitation powers. At the lowest power, the S1 state lies below the bandgap of the left wire and within the bandgap of the right wire. Therefore S1 only couples to propagating states of the first miniband of the left wire. Indeed at the energy $E1$, we observe polariton propagation only in the left wire. In the right wire, the emission is strongly damped. Notice that we also observe polariton propagation at energies different from $E1$ (around $1597.6\, \mathrm{meV}$ and in the second miniband). Polaritons are indeed injected also via the second confined mode in the island, and most probably there is some direct injection of polaritons in the wires on both sides of the island through the spatial tail of the excitation laser beam.

Fig.~\ref{Fig2}(g) shows the spatial intensity profiles measured along the left and right wires at the energy of S1. Exponential spatial decays are measured with very different decay lengths on both sides.
For $P = 5 \,\mathrm{mW}$ [Figs.~\ref{Fig2}(b,e,h)], the opposite situation is observed: S1 lies within a gap of the left wire and propagation occurs in the right one.
At higher power [$P = 7.8 \,\mathrm{mW}$, Fig.~\ref{Fig2}(c)], the island state rises above the bandgap of both wires and polaritons are free to propagate in each wire [see Fig.~\ref{Fig2}(i)].

We can estimate the polariton lifetime from the polariton spatial decay in the wires. In the case depicted in Fig.~\ref{Fig2}(c), at high excitation power, the decay length in both wires is $L_{d}= 35 \,\mu \mathrm{m}$. It is mainly governed by photon leakage through the mirrors. From the slope of the dispersion at the energy of the emission [Fig.~\ref{Fig2}(f)], we can estimate a group velocity of $v_{g}= 2 \,\mu \mathrm{m}/\mathrm{ps}$ and deduce the polariton lifetime $\tau_{pol} = L_{d}/v_{g}= 17.5 \,\mathrm{ps} $. This value is lower by a factor of 2 than expected from the nominal quality factor ($Q \sim 72000$). This could be explained by non-radiative recombination of excitons taking place at the edges of the etched microcavity or by photon scattering induced by disorder and imperfection in the wire gratings. The measured propagation length when S1 lies within a forbidden energy gap is much shorter, of the order of $ 5 \,\mu \mathrm{m}$. This damping is due to destructive interferences within the gap.

We define the Normalized Directionality $ND=\frac{I_{left}-I_{right}}{I_{left}+I_{right}}$, where $I_{left}$ and $I_{right}$ are the integrated intensities measured at energy $E1$ in the left and right wire starting $5\, \mu \mathrm{m}$ away from the island. $ND$ is positive (resp. negative) when polaritons are flowing to the left (resp. right) wire and $ND=0$ when polaritons propagate equally in both wires. Fig.~\ref{Fig2}(k) summarizes the control of the S1 state tunnel coupling into the two wires, showing the variation of $ND$ as a function of the S1 blueshift. This curve is well fitted considering the convolution between the island state linewidth and a step function, reflecting the density of states of the wires observed experimentally.

In the following, we demonstrate the operation of the polariton router. We show how polariton pulses injected in the island can be selectively directed into either of the two wires using a two laser excitation scheme. The non-resonant cw laser beam used previously is now used to control the energy $E1$ of the S1 state with a varying control power $P_{c}$.
A pulsed laser tuned to $1598.6\, \mathrm{meV}$ (which is $0.3\,\,\mathrm{meV}$ above the S1 state energy) and with averaged power $P_{inj}=\,0.1 \,\mathrm{mW}$, is focused onto the island and injects polaritons at $t=0 \,ps$. The $360 \,\mu \mathrm{eV}$ linewidth of the laser pulses determines the spectral range in which we can tune $\mathrm{E}_{1}$ while maintaining an efficient injection of polaritons with the pulsed beam.

\begin{figure}[htbp!]
\centerline{\includegraphics[width=\columnwidth]{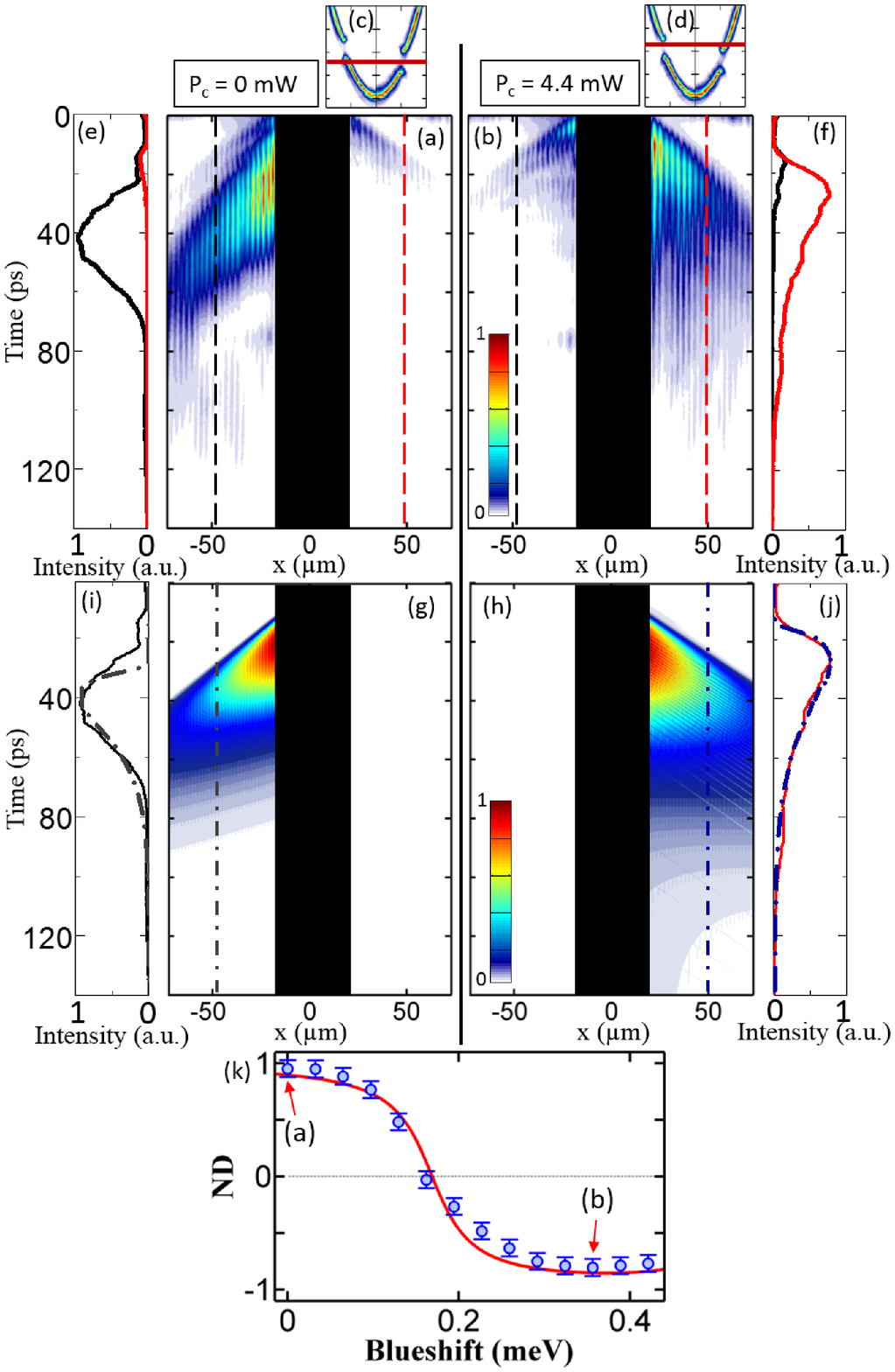}}
\caption{(a-b) Emission of the left and right wire resolved in space and time (a) for $P_c=0\,\mathrm{mW}$ and (b) $P_c=4.4\,\mathrm{mW}$. The emission in the island region is masked (dark region) to remove scattered light from the resonant laser; (c-d) Spectrally resolved far field emission measured on the left  wire (left  part) and on the right wire (right part) for the two considered values of $P_c$. Red lines indicate the energy $E1$ of the S1 state; (e-f) Red (resp. black): Emission intensity measured along the dashed red (resp. black) lines in (c-d) as a function of time (intensity scales are the same for both figures); (g-h) Result of the numerical simulations corresponding to (a) and (b); (i-j) Comparison between simulations (dashed lines) and experiments (solid lines) for the time resolved intensity extracted on the left wire in (a) and (g) and on the right wire in (b) and (h); (k) Measured $ND$ as a function of the S1 blueshift.}
\label{Fig3}
\end{figure}

The time resolved polariton emission measured along left and right wires is shown in Figs.~\ref{Fig3}(a,b). Thanks to the same mechanism as described previously, we observe that at $P_{c}=0 \,\mathrm{mW}$ [Fig.~\ref{Fig3}(a)], polaritons tunnel to the left wire and are prevented from propagation in the right wire. Note that at small times, polaritons propagating with high velocity can be observed in both wires, corresponding to polaritons injected directly into the wire by the spatial tail of the laser. Nevertheless this signal is an order of magnitude weaker.
At $P_{c}=4.4 \,\mathrm{mW}$ [Fig.~\ref{Fig3}(b)], the energy of the island state is blueshifted [Fig.~\ref{Fig3}(d)] and the polariton flow can only propagate in the right wire. Thus using the control laser beam which injects a small exciton population in the island, we are able to route polariton pulses in either of the wires by optical means. Operation of the polariton router is summarized in Fig.~\ref{Fig3}(k), where $ND$ is plotted as a function of the S1 blueshift for this two beam experiment.

Fig.~\ref{Fig3} also presents numerical simulations of the polariton router under such dynamical regime. The simulations are based on a coupled rate equation model accounting for the population dynamics of both S1 and the excitonic reservoir \cite{Wouters2007} (more details about the model are given in the supplementary materials \cite{Supplementary}).
In Fig.~\ref{Fig3}(g) we only examine resonant pulsed injection of polaritons in S1 and their dynamics, considering both radiative recombination within the island and tunneling and propagation into the wires. In Fig.~\ref{Fig3}(h), we add a  population in the excitonic reservoir which not only blueshifts the S1 state but also introduces a dynamical additional polariton population into S1 via stimulated scattering. The overall dynamics is well reproduced, in particular, the temporal profiles measured on each side [see Figs.~\ref{Fig3}(i,j)]. Stimulated relaxation from the reservoir into S1 is responsible for the longer decay observed when the island is blueshifted [Fig.~\ref{Fig3}(j)]. The absolute value of the transmitted beam is linked to this additional polariton population coming from the reservoir, but also to the spectral coupling of the incident pulsed laser beam to the S1 state (which changes when tuning S1), and to the tunnel coupling of S1 to the wire (which also changes when tuning S1). Interplay of these effects result in comparable intensities of the pulses transmitted in the left and right wires.

The operation speed for this polariton device is expected to be limited by the lifetime of the excitonic reservoir used to switch the router between one or the other output wire. This lifetime is typically of the order of a few hundreds of picoseconds \cite{Bajoni2006a}, limiting the operation frequency of the device to several GHz. Faster operation speed could be envisaged using resonant pumping of a higher energy state of the island (in this case the speed would be limited by the polariton lifetime), or by dynamical Stark effect \cite{Hayat2012}. There is an optimal range of pulse width for the polariton router operation. Indeed the pulse spectral width should not exceed that of the energy splitting between S1 and the second confined state in the island (in the present device, this splitting amounts to $1.3 \, \mathrm{meV}$  setting $0.2\, \mathrm{ps}$ as a lower limit for the pulse duration) but should neither be spectrally too narrow so that the pulses overlap properly with S1 when tuning E1. Typically in the present device, the polariton pulse duration can be in the range of $0.2$ to $5\, \mathrm{ps}$.

Finally we are aware that a key issue for real applications is the temperature at which the device is operated. It is limited to cryogenic temperature when GaAs is used because of the instability of excitons in this material at room temperature. Nevertheless impressive progress has been reported concerning the exciton photon strong coupling regime in materials with excitons having higher exciton binding energies. Polariton lasing has been reported at room temperature in large bandgap inorganic materials such as GaN \cite{Christopoulos2007,Daskalakis2013} and ZnO \cite{Li2013}. Also very promising is the case of 2D atomic layers \cite{Mak2010,Liu2014} where strong exciton binding energy \cite{Cheiwchanchamnangij2012} and also strong excitonic interactions \cite{Moody2014} have been measured. We would also like to mention recent results on organic materials \cite{Kena-Cohen2010,Plumhof2014} and in particular the demonstration of strong blueshift induced by exciton interactions \cite{Daskalakis2014}.
  These recent achievements may allow soon implementation at room temperature of polaritonic devices, such as the polariton router for which a proof of principle is demonstrated here.

This work was partly supported by a public grant overseen by the French National Research Agency (ANR) as part of the "Investissements d'Avenir" program (Labex NanoSaclay, reference: ANR-10-LABX-0035), the ANR project Quandyde (Grant No. ANR-11-BS10-001), the French RENATECH network,
and the European Research Council grant Honeypol.

\bibliographystyle{apl}

\begin{thebibliography}{10}
\bibitem{Largeurderaie}
This value of $\Gamma$ is an upper limit: under such non-resonant excitation, spectral broadening occurs due to electrostatic fluctuations in the island induced by the optically injected carriers.
\end{thebibliography}

\end{document}